\documentclass{article}

\usepackage{graphicx}
\usepackage{amsmath}
\usepackage{amssymb}
\usepackage[margin=2cm]{geometry}
\usepackage{natbib}
\usepackage[caption=false]{subfig}

\usepackage{xcolor}

\title{Frequency selectivity of neural circuits with heterogeneous transmission delays}
\author{Akke Mats Houben}
\begin{document}
\maketitle

\begin{abstract}
Neurons are connected to other neurons by axons and dendrites that conduct signals with finite velocities, resulting in delays between the firing of a neuron and the arrival of the resultant impulse at other neurons.
Since delays greatly complicate the analytical treatment and interpretation of models, they are usually neglected or taken to be uniform, leading to a lack in the comprehension of the effects of delays in neural systems.
This paper shows that heterogeneous transmission delays make small groups of neurons respond selectively to inputs with differing frequency spectra. 
By studying a single integrate-and-fire neuron receiving correlated time-shifted inputs, it is shown how the frequency response is linked to both the strengths and delay times of the afferent connections. 
The results show that incorporating delays alters the functioning of neural networks,
and changes the effect that neural connections and synaptic strengths have.
\end{abstract}
\twocolumn

\section{Introduction}
Eventhough the brain is quick to respond, neurons are connected through axons and dendrites that propagate signals with non negligible and replicable delays \citep{Swadlow1985, Swadlow1994}. The transmission delay between two neurons depends on the conduction velocity, related to the diameter of the axon or dendrite \citep{Cullheim1978, Cullheim1979, Gasser1939, Lee1986, Waxman1980} and the properties of the axon and dendrites \citep{Harper1985, Waxman1980}, in combination with the lengh of the path a pulse travels from one neuron to the other ($time = distance/speed$). Conduction delays have been shown to be plastic \citep{Bakkum2008}, indicating that condution delays are tuneable to some extent. On the other hand, the conduction velocities have been found to be activity dependent \citep{Swadlow1974, Thalhammer1994, deCol2008}. Still, given that the response times of neurons to natural stimuli are robust \citep{Mainen1995}, the transmission delay times should be stable over time \citep{Swadlow1985}.

Because of the difficulties involved in the analytical treatment of delays, they are often either neglected or taken to be uniform. This has lead to a lack in the comprehension of the effects of (heterogeneous) delays in neural systems and the absence of delays in principal accounts of the working of neural networks. % (e.g. ...), at times even going as far as the virtual abolishment of time (e.g. Hopfield, rate based models, ...).
Great insights into the functioning and dynamics have been gained by studying networks without (heterogeneous) transmission delays, such as the understanding of the on- and offset of oscillatory activity \citep[e.g.][]{Wang1996, Wilson1972}, signal propagation \citep[e.g.][]{Mehring2003, Reyes2003, Vogels2009}, activity dynamics \citep[e.g.][]{Destexhe2009, Renart2010, vanVreeswijk1996}, and memory storage \citep[e.g.][]{Amit1985, Anderson1972, Brunel2016, Hopfield1982, Klampfl2013}. 

However, there are cases in which the incorporation of transmission delays can lead to drastically different restults. Organisms, and with that their brains, operate in time and as such it is likely there is an importance in the temporal dimension of the activity of the brain. It has already been advocated that transmission delays endow neural networks with much richer dynamics, increasing their functional capacity and possible dynamics \citep[e.g.][]{Chapeau-Blondeau1992, Destexhe1994b, Izhikevich2006, Ostojic2014}, enabling neural communication based on synchrony or spike ordering \cite[e.g.][]{Brette2012, Gautrais1998, Thorpe1990}, and allowing oscillations and synchroniation \citep[e.g.][]{Buzsaki2004, Destexhe1994a, Ernst1995, Geisler2010, Maex2003, vanVreeswijk1994}. 
This paper concerns the connection between transmission delays and the frequencies to which a neuron responds. In particular it exposes the concerted effect of synaptic strenghts and synaptic delays on the frequency selectivity of neurons.

Synaptic connections between (excitatory) neurons are principally understood in light of some variant of a constitutive Hebbian learning process. The resultant view is that for two neurons to be connected by an excitatory connection it means that it is likely to observe a co-occurence of their spiking activity, and the stronger this connection, the more likely they will fire together. 
A synaptic connection can also be understood through the causal effect of a pre-synaptic neuron on the activity of a post-synaptic cell, determined by the synaptic strenght and pre-synaptic neuron type. 
Excitatory pre-synaptic neurons have a depolarising effect and thus promote the post-synaptic firing, in general leading to higher post-synaptic firing rates. Inhibitory neurons lead to hyperpolarisation and thus inhibit or delay post-synaptic firing, leading to reduced post-synaptic firing rates. These effects are more pronounced for stronger connections.

However, by the inclusion of heterogeneous transmission delays one can conceptualise a small circuit of neurons as a cascade of filters with time-delayed and weighted coupling. 
Seen in this way the properties of the afferent connections determine the sub-threshold frequency response of a neuron, with the synaptic weights functioning as feed-forward coefficients in a neural filtering circuit and thus influencing the frequency selectivity of the post-synaptic neuron. 
This means that the effect of synaptic strengths differs from the one given above, and that the effect of a single synapse cannot be completely understood in isolation, but gets a significantly different interpretation when considered in an ensemble of synapses conveing correlated signals.

This paper explores this concept with a single integrate-and-fire neuron receiving correlated and time-shifted inputs. 
It is shown that the sub-threshold frequency response of a neuron is determined by the strengths and relative delay times of the correlated afferent connections. The characteristics of the frequency response of some cases are solved exactly, and qualitative observations are given for more general cases. 
Subsequently, numerical simulations demonstrate the functional significance of this frequency selectivity, the frequency specificity, and show that the described effects hold for a range of input correlations.

\section{Results}
To show the basic principle, consider a leaky integrate-and-fire (IF) neuron, with a membrane potential governed by:
\begin{equation}\label{eq:dv}
    \tau\frac{dv}{dt} + v = G\left(I + \sum_{m=1}^{M} w_m I^{(m)} \right), %\frac{dv}{dt} = \tau^{-1}[G(I + w*I') -v],
\end{equation}
in which $\tau$ is the membrane time constant.
This equation is complemented with a spike-and-reset rule: once the membrane potential $v$ exceeds a threshold $v_{th}$, the membrane potential is directly reset to a reset value $v_r$, after which the membrane potential is again directly governed by (\ref{eq:dv}).

The input is given by the term $G\left(I + \sum_m w_m I^{(m)}\right)$, which denotes the sum of a stochastic input $I$ with correlated and time-shifted inputs $I^{(m)}$, each of which is multiplied by a weight $w_m$.
The input can be interpreted as the output of $M$ similar neurons or groups of neurons that receive strongly correlated input and so produce highly correlated outputs, arriving through connections with different delay times due to the different finite conduction velocities of the different axons and dendrites. The term $G$ is a gain factor which, since the frequency response does not qualitatively depend on the absolute gain but on the relative strenghts of the individual inputs, will be set to be unity in the following analysis of this system.

\begin{figure}
    \includegraphics[width=\columnwidth]{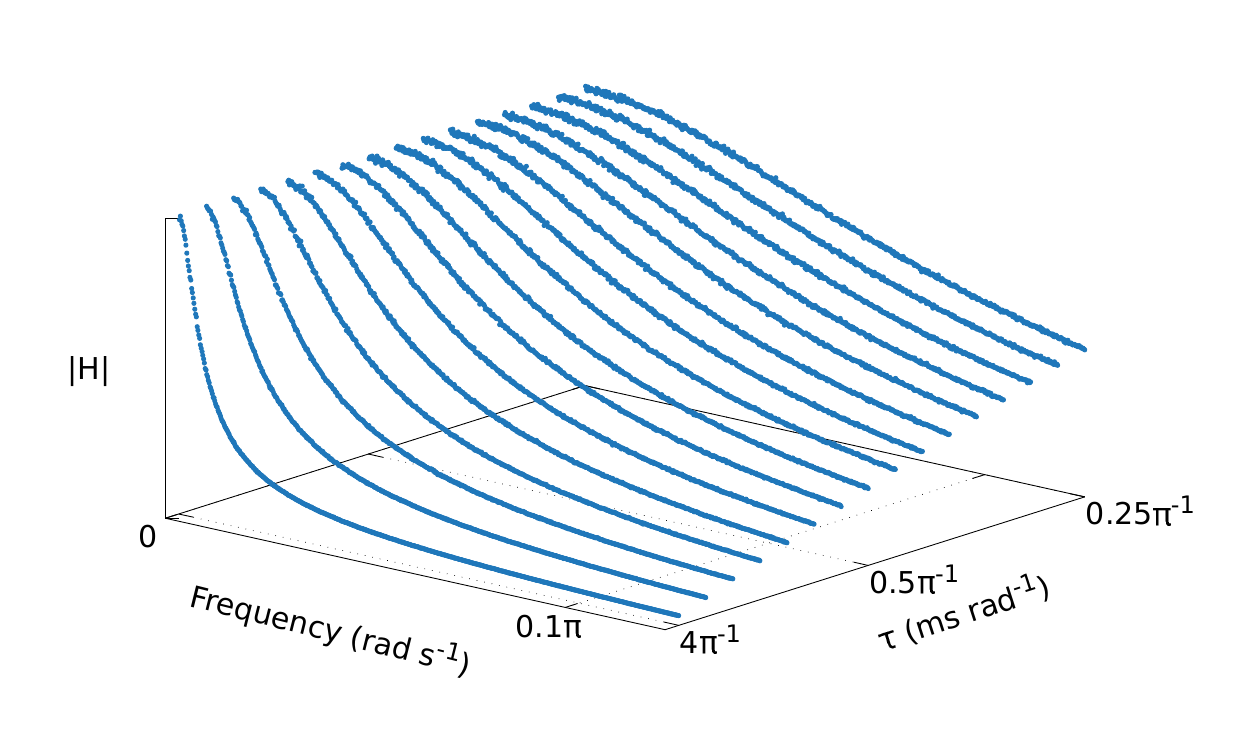}
    \caption{\textbf{Sub-thresold frequency spectrum} of (\ref{eq:dv}) measured for different time-constants $\tau$ (in $s\cdot rad^{-1}$), showing the low-pass filtering effect intrinsic to the membrane equation}\label{fig:lifH} 
\end{figure}

Equation (\ref{eq:dv}) without the spiking mechanism (or with $v_{th} >> \langle I\rangle$) is linear and time-invariant, as such the sub-threshold frequency spectrum of $v$ is given by the product of the intrinsic frequency response $1/(1+\tau s)$, with $s$ being a complex frequency of the form $s=\alpha + \textit{i}\omega$, and the frequency spectrum of the input \citep{Oppenheim1997, Smith2007}:
\begin{equation}\label{eq:lifH}
  \widetilde{v}(s) = \frac{1}{1 + \tau s} \left(\widetilde{I}(s)+\sum w_m\widetilde{I}^{(m)}(s)\right),
\end{equation}
from which we see that the membrane equation has a pole on the real axis at $s=-\tau^{-1}$, corresponding to the characteristic decaying response of the leaky IF neuron \citep{Stein1965, Gluss1967}. Thus the intrinsic frequency response of the membrane acts as a low-pass filter with time constant $\tau$.
Figure \ref{fig:lifH} shows the intrinsic sub-threshold frequency response of the membrane equation (\ref{eq:dv}) for different values of $\tau$.

The frequency transfer function $H(s)$ of linear time-invariant systems can be found by dividing the output spectrum by the input spectrum, $H(s):=\widetilde{V}/\widetilde{I}$ \citep{Oppenheim1997}. Requiring that all the input spike trains are strongly correlated ($\rho =1$) for different positive lag times, $I^{(m)}(t) = I(t-d_m)$ for all $m$, with $d_m>0$, and dividing (\ref{eq:lifH}) by $\widetilde{I}$ leads to the sub-threshold transfer function of the membrane equation (\ref{eq:dv}):
\begin{equation}\label{eq:H}
  H(s) = \frac{\left(1+\sum w_m {e}^{-d_ms}\right)}{1+\tau s}.% + \sum\left[\frac{b_m}{1+tau_m}\right]}
\end{equation}

\subsection{Frequency response}\label{sec:poze}
The frequency response of the neuronal circuit can be understood by finding the roots of the denominator and the nominator of the transfer function (\ref{eq:H}), which will respectively give the positions of the poles and zeros of the frequency response.
Since we are not occupied with feedback components other than the intrinsic membrane dynamics, (\ref{eq:H}) has just a single pole which position is, as shown above, determined by the membrane time-constant and lies at $p=-1/\tau$. The rate of decay of the membrane potential is proportional to the distance of this pole to the origin, which can also be seen from the homogeneous solution to (\ref{eq:dv}). 

The zeros of (\ref{eq:H}) correspond to frequencies at which $H(s)$ vanishes, corresponding to the input frequencies to which the neuron responds minimally, so the roots of the numerator correspond to frequencies that are attenuated by the circuit. Since $d_m \in \mathbb{R}$ it is difficult to find exact expressions for the roots of (\ref{eq:H}). 
Exact analysis of the roots in general cases is thus beyond the scope of this paper, but it is straightforward to plot (\ref{eq:H}) in order to gain a qualitative insight into the frequency response. 

However it will be instructive to treat some cases in which the roots can be obtained analytically. For the folowing it is assumed that each delay time $d_m$ is an integer multiplying some basic time unit $\delta_0 = 1/(2\pi f_{max})$ in $sec \cdot rad^{-1}$. The highest frequency to be analysed is determined by $f_{max}$, or conversely $f_{max}$ can be determined by requirement on $\delta_0$ to fit the desired values of $d_m$. In the following a normalised frequency will be used ($f_{max}$ will be normalised to unity) such that $\delta_0 = 1/(2\pi)$ $sec\cdot rad^{-1}$.
Two specific cases will be treated analytically: a neuron receiving one additional input ($M=2$) for a delay time $d>0$, resulting in a `comb filtering', and the case of a neuron receiving 2 additional inputs ($M=3$) with $d_2 = 2d_1, d_1>0$ such that the nominator can be transformed into a quadratic polynomial. Some of the results will be extended to the more general case in which $d_2 = nd_1$, for $n\in\mathbb{N}$. Afterwards some qualitative observations about the zeros of (\ref{eq:H}) will be made.
\\

\begin{figure*}
    \centering
    \subfloat[\label{fig:schematicComb}]{\includegraphics[width=.2\textwidth]{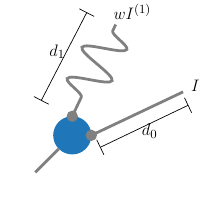}} % schematic
    \subfloat[\label{fig:resComb1}]{\includegraphics[width=.35\textwidth]{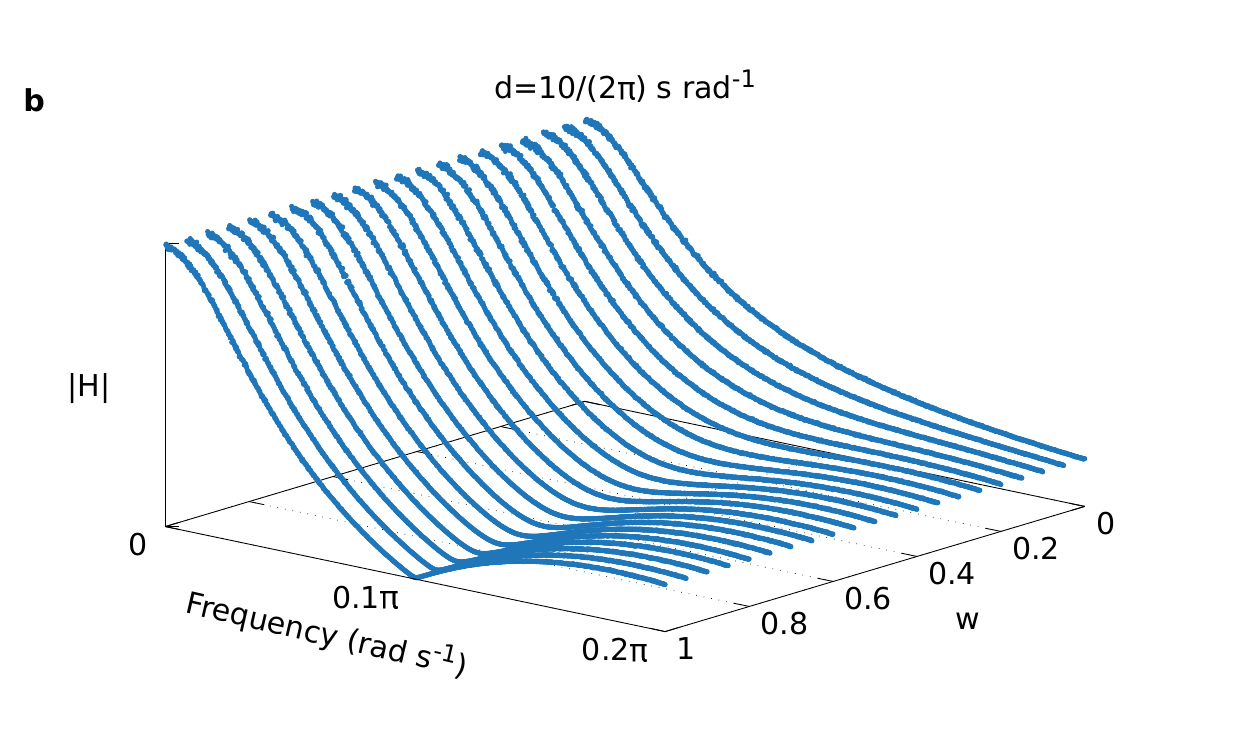}}
    \subfloat[\label{fig:resComb2}]{\includegraphics[width=.35\textwidth]{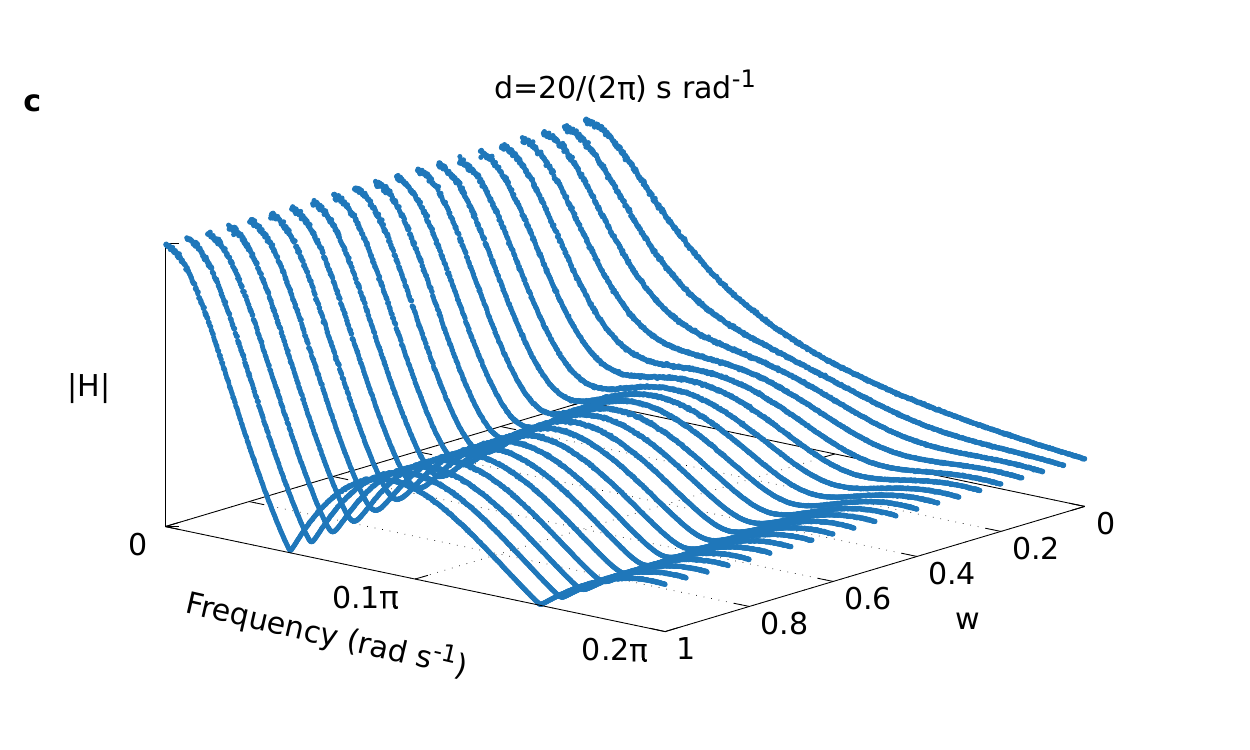}}
    \caption{\textbf{`Comb'-filtering}: a) schematic drawing indicating neuron wiring receiving 2 inputs with differing delay times $d_0<d_1$, due to differing path lengths; b-c) measured sub-threshold frequency spectra for different synaptic weight $w$ values, for two different delay times}\label{fig:comb}
\end{figure*}

%\subsubsection{Analysis of zeros}
In the case of $M=2$ the nominator of (\ref{eq:H}) has periodically distributed roots
\begin{align}\label{eq:onez_rad}
  z_n = &-w^{\frac{1}{d}}e^{\frac{\mathit{i}(2n+k)\pi}{d}},\\
    &\text{for $n = 1, 2, 3, ...$ and }\nonumber \\ 
    & k = \begin{cases} 
1, &\text{$d$ even} \\
0, &\text{$d$ odd,}
\end{cases} \nonumber
\end{align}
from which we immediately see that the attenuated frequencies (given by the the angles $\angle z_n$) are determined by the delay time $d$, and that the weight $w$ only influences the amount of attenuation (given by the magnitudes $|z_n|$). Figure \ref{fig:comb} shows the frequency responses for some different values for $d$, also demonstrating the reason for the conventionally used name `comb filter'.

\begin{figure*}
    \centering
    \subfloat[\label{fig:schematicQuad}]{\includegraphics[width=.2\textwidth]{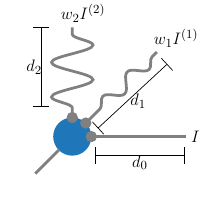}} % schematic
    \subfloat[\label{fig:resQuad1}]{\includegraphics[width=.35\textwidth]{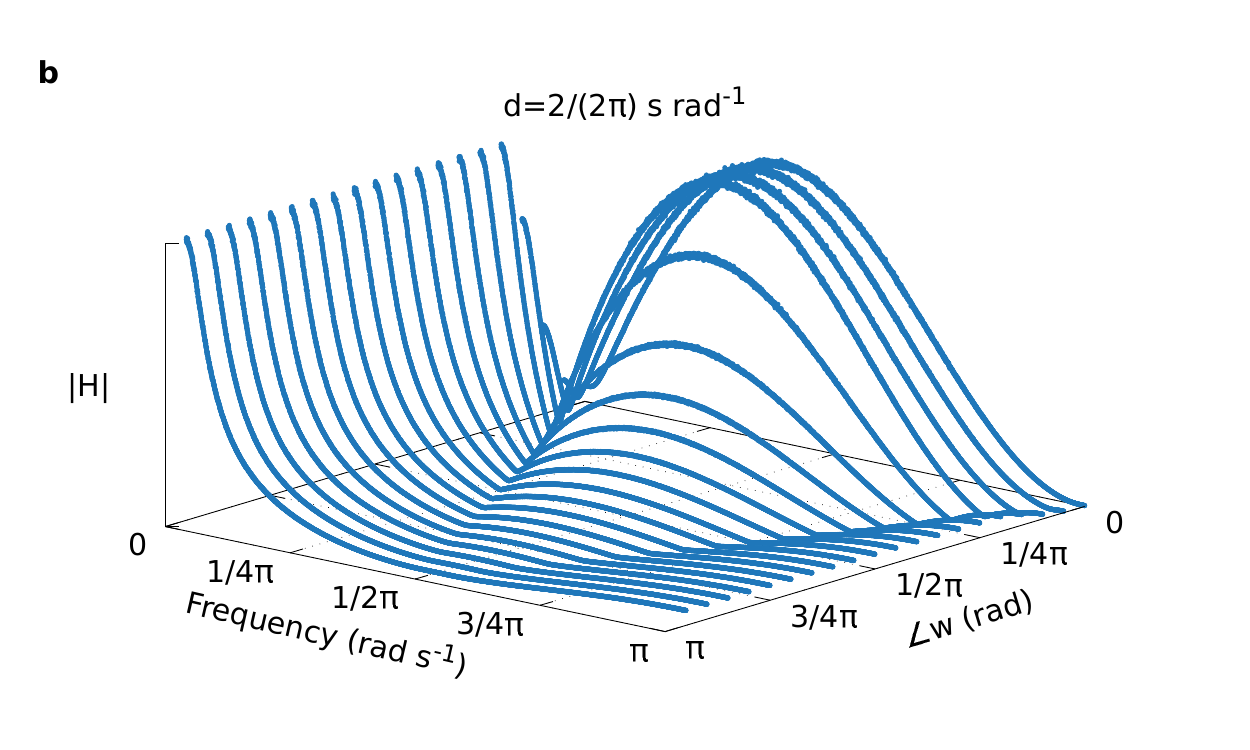}}
    \subfloat[\label{fig:resQuad2}]{\includegraphics[width=.35\textwidth]{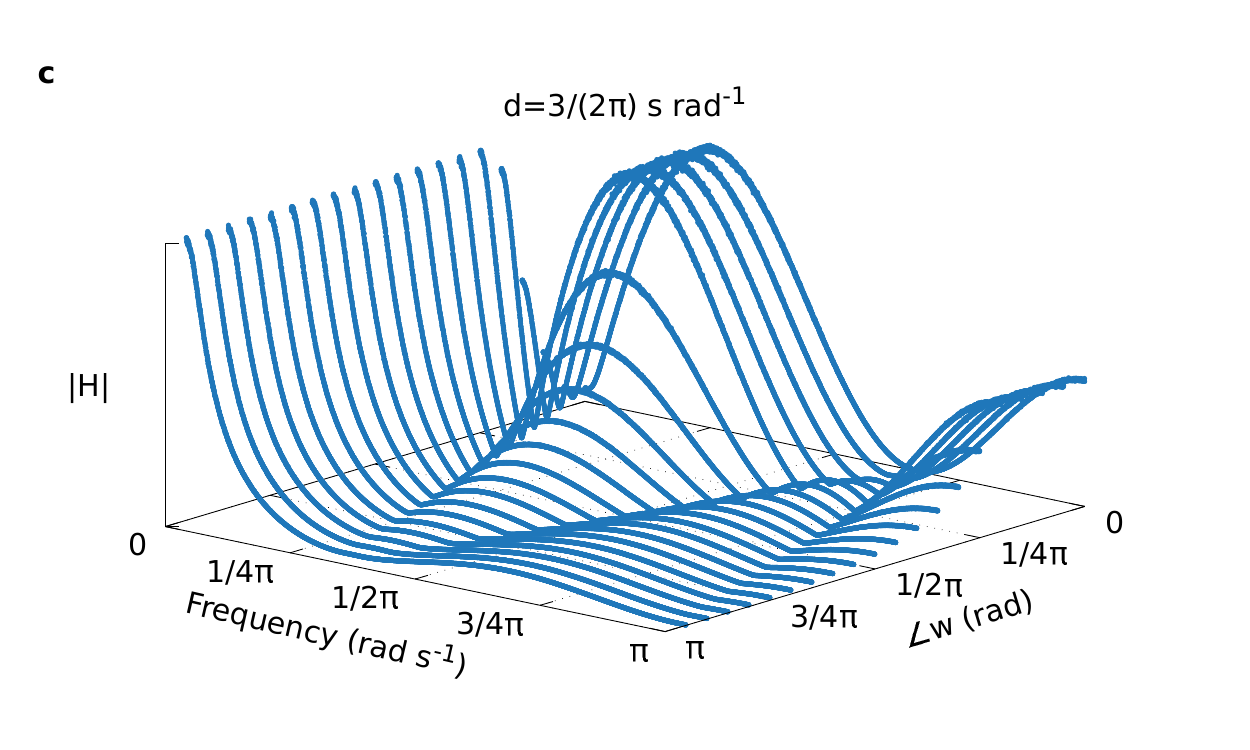}}
    \caption{\textbf{`Quadratic' filters}: a) schematic of neuron wiring; b-c) measured sub-threshold frequency spectra for neurons receiving 3 inputs with $w_2=1$ and different values for $w_1$, for two different delay times $d_1$, and $d_2=2d_1$}\label{fig:quad}
\end{figure*}

A more interesting case is to analyse the zeros of the sub-threshold transfer function
\begin{equation}\label{eq:ex1}
  H(s) = \frac{\left(1+w_1e^{-ds}+w_2e^{-2ds}\right)}{1+\tau s},
\end{equation}
describing a neuron receiving $M=3$ inputs with harmonically related delays (see figure \ref{fig:schematicQuad}), which can be obtained exactly by defining $\sigma(s) = e^{ds}$ and multiplying the transfer function by $\frac{\sigma^2}{\sigma^2}$. This allows to rewrite the nominator $N(s)$ of the transfer function as
\begin{equation}
  N(\sigma) = \left(\sigma^2 + w_1\sigma + w_2 \right), \nonumber
\end{equation}
which is quadratic in $\sigma$. Now any standard strategy to obtain the roots of $N(\sigma)$ can be employed, leading to:
\begin{equation}
  \sigma = \frac{-w_1 \pm \sqrt{w_1^2 - 4w_2}}{2}. \nonumber
\end{equation}
Substituting back $e^{ds}$ for $\sigma$, we can use a similar formula as in the case of $M=2$, leading to
\begin{equation}\label{eq:ex1_zeros}
  z_n = -\left(\frac{-w_1 \pm \sqrt{w_1^2 -4w_2}}{2}\right)^{\frac{1}{d}} e^{\frac{\mathit{i}(2n+1-k)\pi}{d}}, \nonumber
%\text{ for $d$ in $s\cdot rad^{-1}$},%\\
  %z_k &= -\left(\frac{-w_1 \pm \sqrt{w_1^2 -4w_2}}{2}\right)^{\frac{1}{d}} e^{\frac{\mathit{i}\pi(2k+\kappa)}{d}}, \text{ for $d$ in $s$},
\end{equation}
with $n$ and $k$ the same as in (\ref{eq:onez_rad}), 
showing that the zeros of (\ref{eq:ex1}) repeat with a period $1/d_1$, and that each of these intervals contains 2 zeros. In this case the frequencies at which the zeros are positioned are influenced by both the delay times and the connection weights. Notice that since each $w_m$ is required to be real, the complex zeros occur in conjugate pairs. Figures \ref{fig:resQuad1} and \ref{fig:resQuad2} show the frequency responses of for some values for $d$ and $w$.

The periodicity of the zeros extends also to a more general case where $d_2$ is an integer multiple of $d_1$. In this case the zeros repeat again with period $1/d_1$, but now each interval contains $d_2/d_1$ zeros.

As said before, in general it will be difficult to determine the zeros of (\ref{eq:H}) exactly. Still some of the above observations can be extended by graphing the magnitude of $H(s)$ with $s=\mathit{i}\omega$ with respect to $\omega$. The first observation is that the number of zeros in the $[0, \pi)$ interval is determined by the longest delay time. More specifically: longer delay times lead to a more rapid succession of zeros. Indeed, this can be inferred from the exact treatment in the last section, in which higher order numerators lead to a faster sucession of zeros. The second observation is that when the delays have a harmonic relationship, with each $d_m$ being an integer multiple of $d_1$, the pattern of zeros occurs periodically with a period of $1/d_1$. Thirdly, in case of complex roots, these roots have to occur in conjugate pairs, since the weights are defined to be real. Finally, the period of repetition of zeros is determined by the delay times, but crucially in case of complex roots, the weights determine the exact attenuated frequency within each interval. This leads to the important observation that synaptic plasticity not only alters the susceptibility, and with that the frequency, of post-synaptic firing, but alters the frequency selectivity of the neuronal circuit.

\subsection{Discrimination of inputs}\label{sec:twoneurons}
\begin{figure*}
  \subfloat[\label{fig:res2n_schematic}]{\includegraphics[width=.33\textwidth]{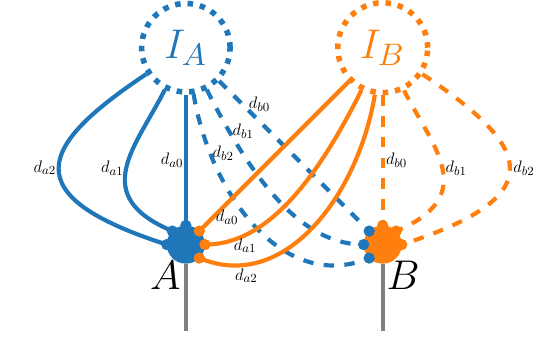}}
  \subfloat[\label{fig:res2n_boxplot}]{\includegraphics[width=.33\textwidth]{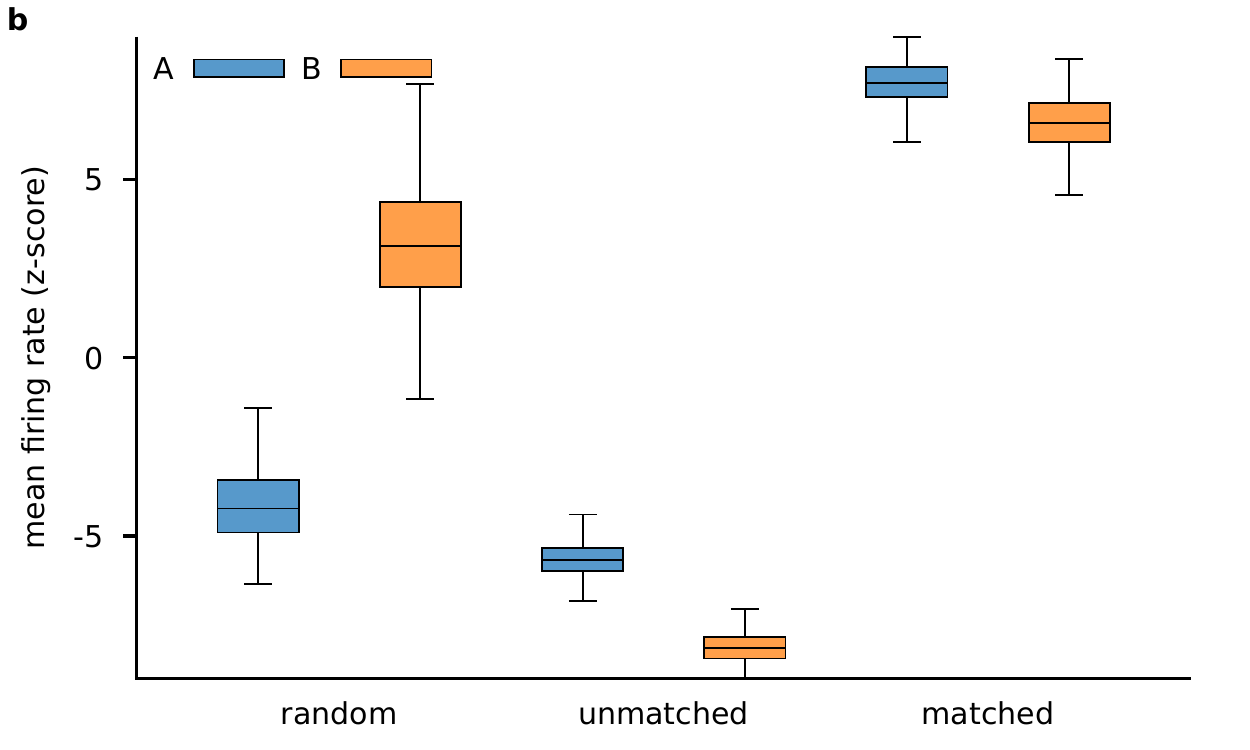}}
  \subfloat[\label{fig:res2n_time}]{\includegraphics[width=.33\textwidth]{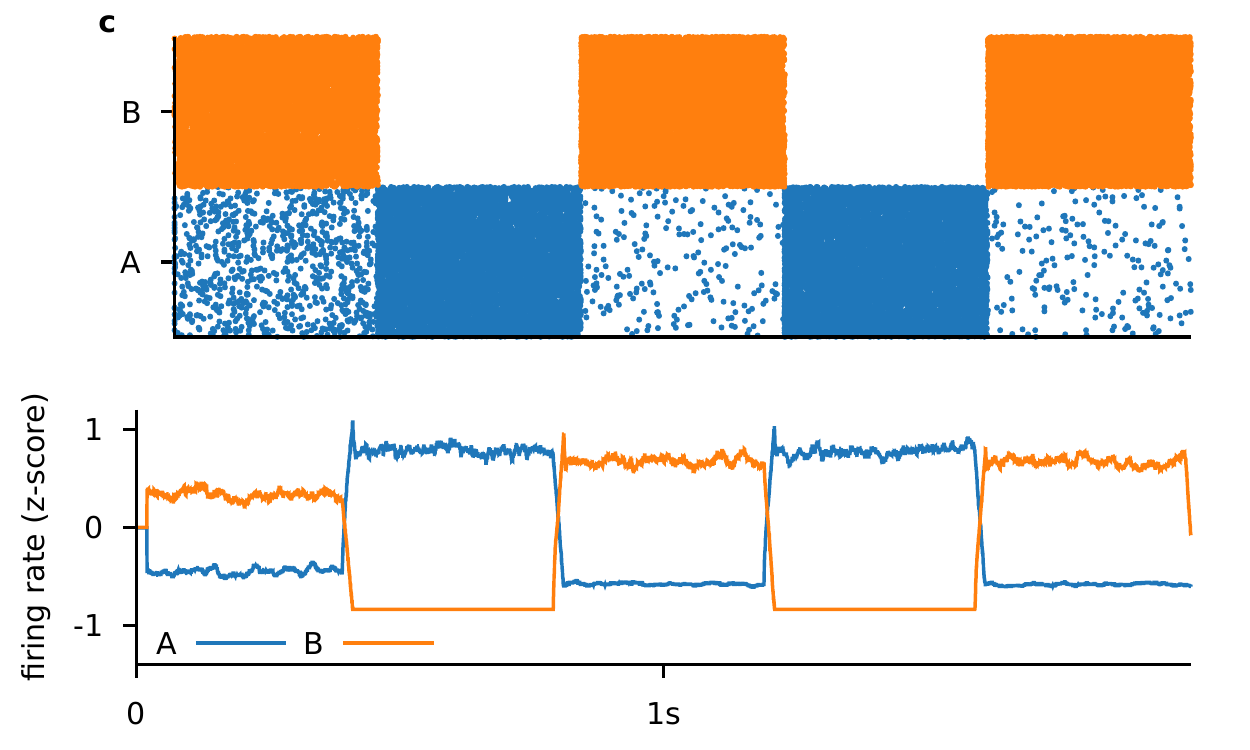}}
\caption{\textbf{Frequency selectivity}: a) schematic indicating neuron wiring. Both neurons receive input both from a source matched to their own sub-threshold frequency selectivity as well as from a source matched to the other neuron. One neuron receives, through an identical set of $M$ input-lines, both the matched and unmatched source. In the figure neuron A receives $M=3$ inputs (indicated by the solid blue input lines) from source $I_A$ with three different delay-times ($d_{a0}<d_{a1}<d_{a2}$) and weights ($w_{a0}$, $w_{a1}$ and $w_{a2}$), and through the same synapses the input of source $I_B$ (indicated by the solid orange input lines). Neuron B receives input from source $I_A$ (dashed blue input lines) and $I_B$ (dashed orange input lines) with the same connection parameters between sources, but different from those of neuron A.
b) box-plots indicating difference in firing rates (transformed into z-scores) of each neuron during white noise (left 2 box-plots), A-matched (middle box-plots) and B-matched (right most plots) input. c) spike raster and averaged spike-rates of neuron A and B in response to a white-noise input (white area), A-matched input (blue areas) and B-matched input (orange areas). Top plot shows the spike timings of $500$ repetitions of the same trial. Bottom plot indicates the average firing rate (transformed to z-scores) of the $500$ repetitions per neuron.
}\label{fig:res2n}
\end{figure*}

The filtering capabilities endowed by heterogeneous transmission delays are not purely theoretic, but can be shown to have a definite effect for the functioning of neural circuits. 
Driving two neurons with differing frequency selectivity (see methods section for the neuron parameters) with a communal input that consists of a white noise during the first $200ms$ and subsequently alternates between two filtered white noise signals, each matching the frequency response of one of the neurons, shows that neurons respond selectively to their matched input (see figure \ref{fig:res2n}).
Whereas both neurons respond with low firing rates during the white-noise input interval (on average $0.0$ ($\sigma=0.00$) spikes/second for neuron A and $0.04$ ($\sigma=0.01$) spikes/second for neuron B. During the matched input intervals a clear distinction between the outputs of the two neurons is observed ($36.75$ ($\sigma=3.72$) spikes/second for neuron A versus $0.03$ ($\sigma=0.20$) spikes/second for neuron B during A-matched input ($t(998)=468.00, p=0.00$), $1.03$ ($\sigma=1.13$) spikes/second for neuron A versus $53.66$ ($\sigma=5.12$) spikes/second for neuron B during B-matched input ($t(998)=-292.70, p=0.00$)). 

Thus neurons are sensitive to their matched spectral input and can selectively respond to differing spectral inputs: each neuron responded less during the other neuron matched input with respect to the white noise input ($t(499)=-175.06, p=0.00$ for neuron A, $t(499)=-32.42, p=6.89e-125$ for neuron B). In order to visualise the specificty of neuronal filtering circuits, neuronal circuits with sub-threshold frequency responses which only pass a narrow band of frequencies are stimulated with narrow band filtered white noise input with different center band frequencies (see methods for the neuron parameters). 
\begin{figure}
  \includegraphics[width=\columnwidth]{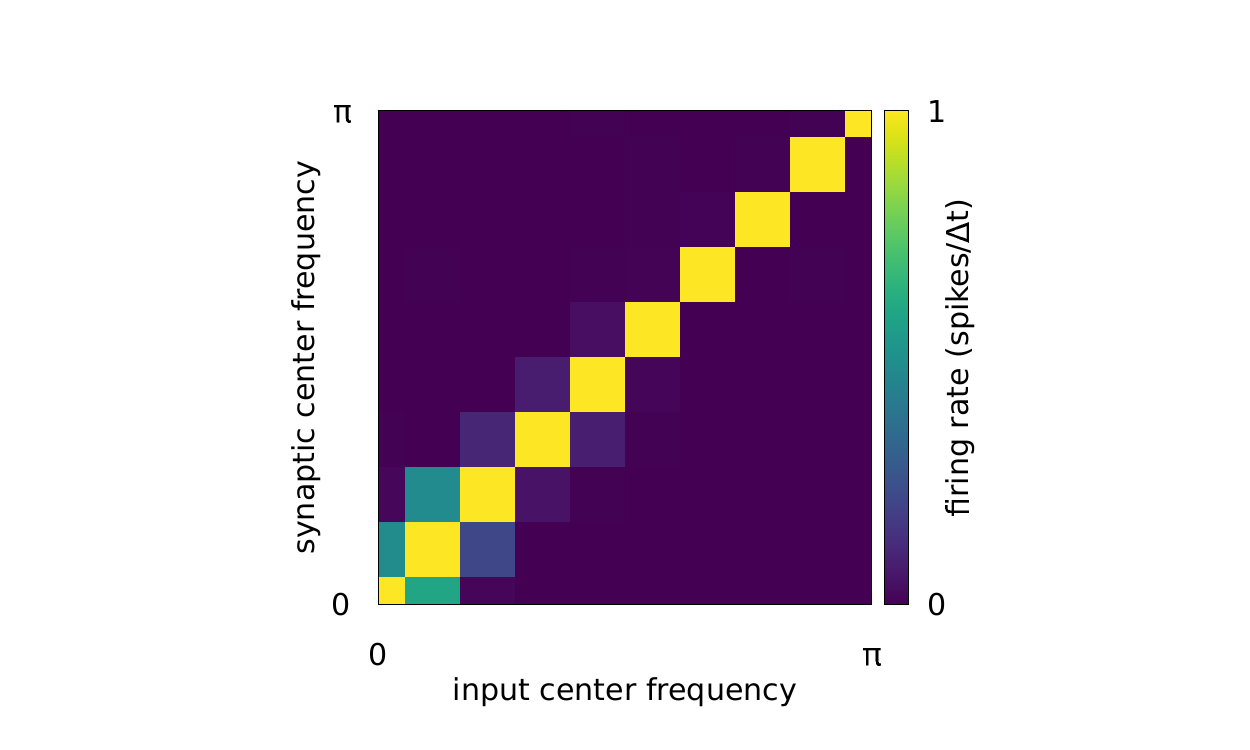}
  \caption{\textbf{Specificity of neuron response}: output firing rate of neurons with narrow band-pass frequency selectivity ($M=13$), with different center-frequencies (x-axis), in response to band-pass filtered white noise input with different center-frequencies (y-axis). Diagonal entries correspond to matched input and synaptic-filtering center-frequency. Off-diagonal entries correspond to increasing disparity between the synaptic-filter and input center-frequencies}\label{fig:res2_specificity}
\end{figure}
Figure \ref{fig:res2_specificity} shows that each neuron responds primarily to their matched input, examplified by the maximal response values in the diagonal entries, and that the responses diminish rapidly with different center frequencies, as shown by a rapid diminishing of activity in the off-diagonal entries. Neuronal circuits with sepecific wiring are thus capable to respond strongly to matching frequency inputs, while supressing their input to non-frequency matched inputs. In order to suppress most of the frequency spectrum and only pass an increasingly narrow band, a neuron needs to receive more and more inputs. However the neurons simulated for figure \ref{fig:res2_specificity} received only $13$ inputs, much less than the estimates of the average number of inputs received by neurons.

\subsection{Robustness for $\rho \neq 1$}
Throughout this paper it was assumed that the inputs the post-synaptic neuron receives through the different synapses were perfectly correlated $\rho=1$. Eventhough the intrinsic noise levels of neurons are low \citep{Mainen1995}, the great number and diversity of synaptic inputs still likely leads to each neuron being subjected to a `unique' noise source. Indeed correlations in the output of any two neurons are generally weak \citep{Cohen2011}. Perfect correlation is thus, in general, unlikely.

\begin{figure}
    \centering
    \subfloat[\label{fig:res2_plotH}]{\includegraphics[width=0.5\columnwidth]{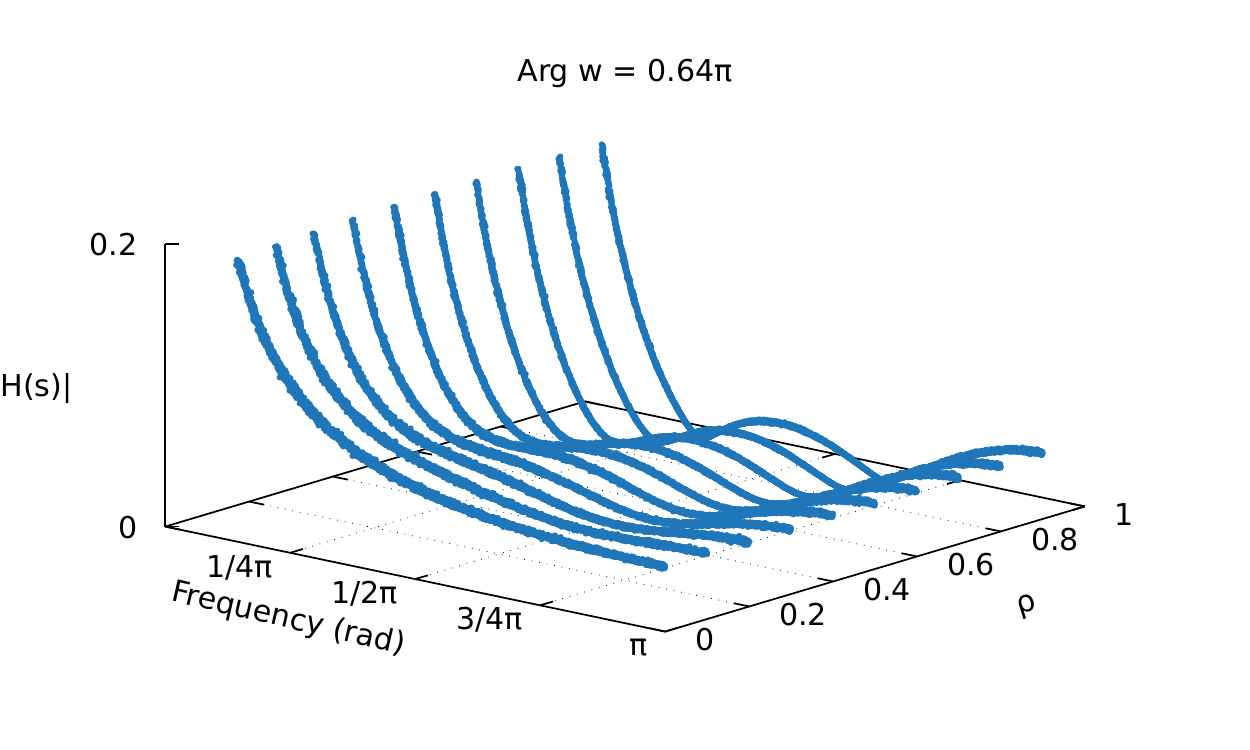}}
    \subfloat[\label{fig:rho2neurons_boxplot}]{\includegraphics[width=.5\columnwidth]{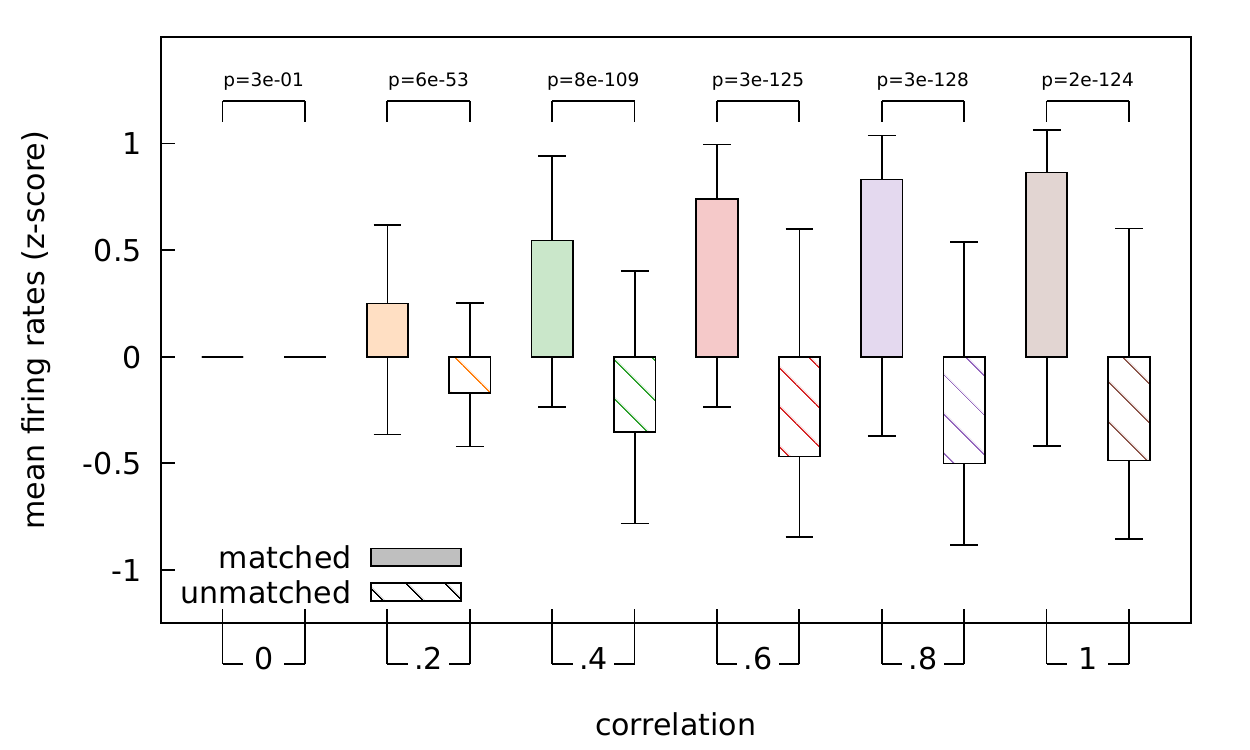}}

    \subfloat[\label{fig:rho2neurons_time}]{\includegraphics[width=\columnwidth]{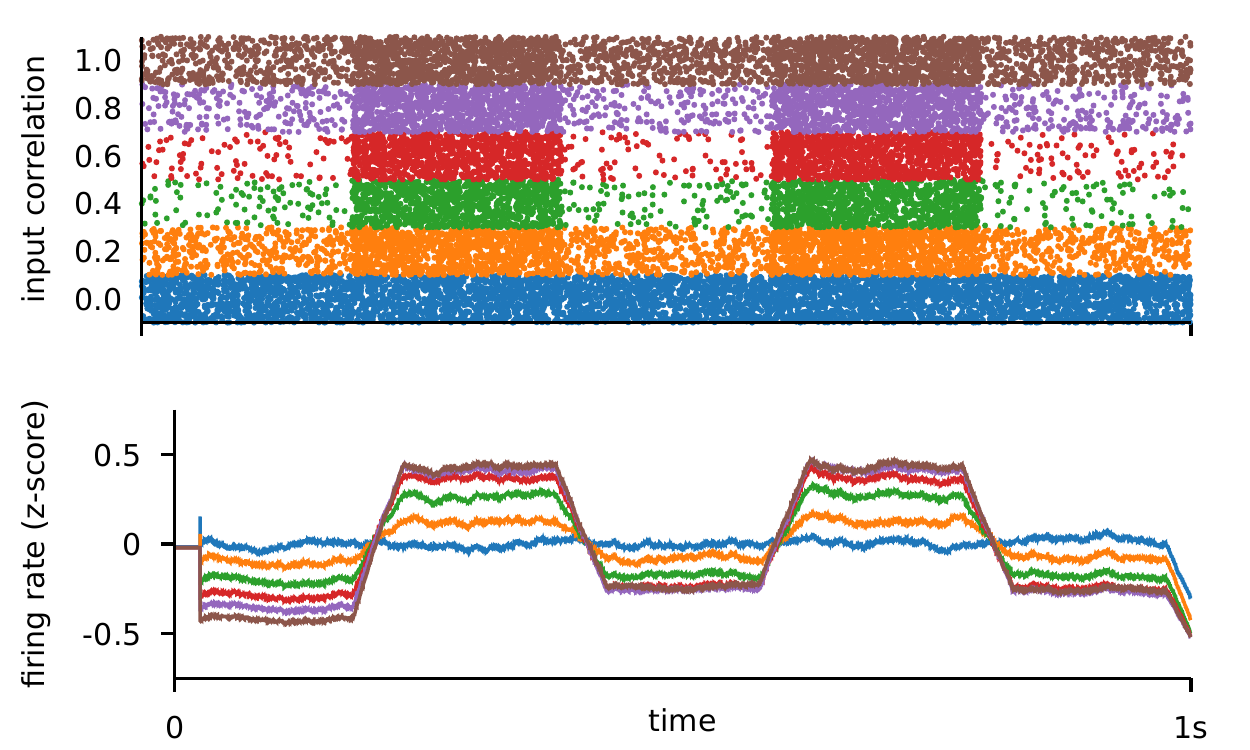}}
    \caption{
\textbf{Robustness to de-correlated inputs}: a) measured sub-threshold frequency spectra for decreasing inter-synapse input correlation $\rho$, showing fast degradation of frequency selectivity with decreasing correlation. 
b) boxplot indicating z-scores of firing rate in response to matched (solid boxes) and unmatched (dashed boxes) with different inter-input correlations (x-axis). 
c) Spike raster and averaged firing rates of $1000$ neurons with random synapse parameters (number and position of conjugate zeros), in response to input alternating between matched and unmatched input, repeated for different inter-synapse input correlations (indicated with different colors, corresponding to the colors of panel b). Top plot indicating the spike timings of the neurons for 6 different correlation values (indicated with different colours). Bottom plot indicates the average of the firing rates, transformed into z-scores, of all the neurons for each inter-synapse correlation value.
}\label{fig:res2}
\end{figure}

In order to investigate the tolerance to non-perfectly correlated inputs, a neuron is simulated receiving 2 additional inputs $M=3$, with different correlation between the inputs arriving through the different delay lines, thus 
\begin{equation}\label{eq:rhonoise}
  I^{(m)} = \rho I + (1-\rho) \eta_m,
\end{equation}
with $\eta_m$ being a similar noise source as $I$. Fourier transforming the sub-threshold responses shows, as visible in figure \ref{fig:res2_plotH}, that with decreasing correlation the shape of the `measured' sub-threshold frequency spectrum quickly reduces to that of the intrinsic sub-threshold response of the membrane equation (equal to the spectrum on the left for $\rho=0$). This suggests that even relatively small disrelations between the inputs abolishes the frequency selectivity of a neuron described in this paper. 

However, again simulating the two neurons adapted to discriminate between two spectrally different inputs, but this time with each neuron receiving white noise input with strength $(1-\rho)$ in addition to the spectrally shaped input (with strength $\rho$), leads to a suprising finding. Figures \ref{fig:rho2neurons_boxplot} and \ref{fig:rho2neurons_time} show the results as in figures \ref{fig:res2n_boxplot} and \ref{fig:res2n_time}, but now for different values of input correlation.
These results show that \emph{functionally} the frequency selective effect due to the input parameters is still present for correlations lower ($\rho=.2$) than for which the frequency response shaping effect is visually prominent from the sub-threshold frequency spectrum (c.f. figure \ref{fig:res2_plotH}).

Thus importantly, eventhough from the `measured' frequency spectra the effect of the inputs on the frequency selectivity seemed to be negligible, the frequency selectivity of the neurons are \emph{functionally} still significantly shaped by the input parameters, and this frequency selectivity is predictable by the theory presented in this paper.

\section{Discussion}
This paper shows that delays in the transmission of signals between neurons have a determinate effect on the frequency response of small neural circuits, making it possible for neural networks to act as finite impulse response filters. It is shown that the length of the delay time but importantly also the strengths of the connections determine the frequency selectivity, and that the frequency response of a neuron due to its afferent connections can be characterised by the analysis of the strengths and delay times of the incoming connections.
Numerical simulations demonstrated that neural networks can be constructed in which neurons with differing connections from overlapping input sources can differentially respond to spectrally different inputs, and can do so with high specificity, thus opening up the frequency domain for usage in neural communication. Finally, it is found that with diminishing input correlation the measured sub-threshold frequency spectra do not show clear signs of the effects of the input parameters, but the frequency selectivity is still functionally present.

The idea of filtering by neurons is not new: a \emph{receptive field} is essentially a filter. It is however important to note that the frequency filtering as treated in this paper is distinct from a filtering of information (as a receptive field does). The filtering described here is a power filtering: the attenuation (and accentuation) of the magnitudes of certain frequencies, leading to differential responses to inputs with different frequencies, which does not necessarily alter the signal-to-noise ratio \citep{Lindner2014}, but alters the dynamics of signal transmission.
Thus, unless spectral content directly conveys information, the effect of delays as presented in this paper does not constitute a `neural code' as such. Rather, it plays an indirect role by making it possible to dynamically route signals (for example by amplifying certain frequency bands; excitation or inhibition of other neurons by the frequency selective neuron or; resetting the phase of oscillatory populations). Through these effects neural codes can be transmitted, so this type of filtering constitutes a medium, rather than the message.

\subsection{Reinterpretation of synaptic strengths}
The presented results show that with the inclusion of transmission delays the effect of synaptic connectivity is different from the generally accepted interpretation: the strenght of a synaptic connection determines not merely the frequency of post-synaptic firing, but co-determines the frequencies to which the post-synaptic neuron responds. This sub-threshold frequency response is determined by the full ensemble of synapses transmitting correlated inputs to a neuron, thus the effect of the strength of a single synapse on the activity of the post-synaptic neuron cannot be understood in isolation, but only in relation to the other synapses. In the light of this observation, the interpretation of the meaning of synaptic strength and the role of synaptic plasticity might need to be reconsidered. It will be crucial, of course, to test the predictions of the theory presented in this paper experimentally. 

\subsection{Interpretation of the inputs}
During the theoretical treatment of the frequency selectivity due to synaptic inputs the time-shifted inputs were taken to be perfectly correlated. Eventhough it was shown that perfect correlation was not needed for neural circuits to retain functionally their frequency selectivity, the correlation levels for which the frequency filtering was qualitatively prominent are higher than reported correlations between pairs of neurons \citep{Cohen2011}. This raises the question how to interpret the inputs used in this study. 

A first posibility, which is also the most simple explanation, is to let the different delayed inputs emerge from the same, or largely overlapping sources, through different transmission lines. In this way the unique noise sources are reduced to that due to synaptic transmission and the propagation of signals along axons and dendrites. It is, by the knowledge of the author, not known whether such connection patterns exist in the brain, but in any case this solution would require a very specific wiring of neural circuits. 

Another option arises by observing that the results of the numerical situations show (section \ref{sec:twoneurons}) that switching the input from a unmatched to a matched input elicits a rapid response in the matched neuron, as visible from the switches between the different inputs in figure \ref{fig:rho2neurons}. From the viewpoint of a neuron this is equivalent to its input switching from uncorrelated (with arbitrary spectral content) signals, to correlated (and frequency matched) signals. Thus neurons need only transient pre-synaptic synchronisation to detect their matched spectral input. 

As a third explaination, eventhough single cell to single cell correlations are low, and seem to actively be kept low even when driven by the same input \citep{Renart2010, Graupner2013}, collectively coherent activity often co-occurs with irregular firing in single neurons \citep{Buzsaki2012}. Network level oscillatory activity can arise from sparsely interconnected and weakly correlated neurons \citep{Brunel1999, Brunel2000}, showing that the correlation between pairs of neurons can be low contemporary with the pooled activation of sub-sets of a population showing stronger correlations. Thus the different inputs can be correlated to a high degree if we interpret the input lines of (\ref{eq:dv}) as each receiving the pooled activity from a population of (sparsely) interconnected and (weakly) correlated neurons.

\section{Acknowledgements}
%The author would like to express special thanks to Pieter Suurmond for the inspiring classes and discussions on filters and systems which have been a great inspiration for the presented research.

\bibliographystyle{apalike}
\bibliography{article2.bbl}%{library.bib}

\begin{thebibliography}{}

\bibitem[Amit et~al., 1985]{Amit1985}
Amit, D.~J., Gutfreund, H., and Sompolinsky, H. (1985).
\newblock Spin-glass models of neural networks.
\newblock {\em Physical Review A}, 32(2):1007.

\bibitem[Anderson, 1972]{Anderson1972}
Anderson, J.~A. (1972).
\newblock A simple neural network generating an interactive memory.
\newblock {\em Mathematical biosciences}, 14(3-4):197--220.

\bibitem[Bakkum et~al., 2008]{Bakkum2008}
Bakkum, D.~J., Chao, Z.~C., and Potter, S.~M. (2008).
\newblock Long-term activity-dependent plasticity of action potential
  propagation delay and amplitude in cortical networks.
\newblock {\em PLOS one}, 3(5):e2088.

\bibitem[Brette, 2012]{Brette2012}
Brette, R. (2012).
\newblock Computing with neural synchrony.
\newblock {\em PLoS Computational Biology}, 8(6).

\bibitem[Brunel, 2000]{Brunel2000}
Brunel, N. (2000).
\newblock Dynamics of sparsely connected networks of excitatory and inhibitory
  spiking neurons.
\newblock {\em Journal of Computational Neuroscience}, 8(3):183--208.

\bibitem[Brunel, 2016]{Brunel2016}
Brunel, N. (2016).
\newblock Is cortical connectivity optimized for storing information?
\newblock {\em Nature neuroscience}, 19(5):749--755.

\bibitem[Brunel and Hakim, 1999]{Brunel1999}
Brunel, N. and Hakim, V. (1999).
\newblock Fast global oscillations in networks of integrate-and-fire neurons
  with low firing rates.
\newblock {\em Neural Computation}, 11(7):1621--1671.

\bibitem[Buzs{\'a}ki and Draguhn, 2004]{Buzsaki2004}
Buzs{\'a}ki, G. and Draguhn, A. (2004).
\newblock Neuronal oscillations in cortical networks.
\newblock {\em Science}, 304(5679):1926--1929.

\bibitem[Buzs{\'a}ki and Wang, 2012]{Buzsaki2012}
Buzs{\'a}ki, G. and Wang, X.-J. (2012).
\newblock Mechanisms of gamma oscillations.
\newblock {\em Annual review of neuroscience}, 35:203--225.

\bibitem[Chapeau-Blondeau and Chauvet, 1992]{Chapeau-Blondeau1992}
Chapeau-Blondeau, F. and Chauvet, G. (1992).
\newblock Stable, oscillatory, and chaotic regimes in the dynamics of small
  neural networks with delay.
\newblock {\em Neural Networks}, 5(5):735--743.

\bibitem[Cohen and Kohn, 2011]{Cohen2011}
Cohen, M.~R. and Kohn, A. (2011).
\newblock Measuring and interpreting neuronal correlations.
\newblock {\em Nature Neuroscience}, 14(7):811.

\bibitem[Cullheim, 1978]{Cullheim1978}
Cullheim, S. (1978).
\newblock Relations between cell body size, axon diameter and axon conduction
  velocity of cat sciatic $\alpha$-motoneurons stained with horseradish
  peroxidase.
\newblock {\em Neuroscience letters}, 8(1):17--20.

\bibitem[Cullheim and Ulfhake, 1979]{Cullheim1979}
Cullheim, S. and Ulfhake, B. (1979).
\newblock Relations between cell body size, axon diameter and axon conduction
  velocity of triceps surae alpha motoneurons during the postnatal development
  in the cat.
\newblock {\em Journal of Comparative Neurology}, 188(4):679--686.

\bibitem[De~Col et~al., 2008]{deCol2008}
De~Col, R., Messlinger, K., and Carr, R.~W. (2008).
\newblock Conduction velocity is regulated by sodium channel inactivation in
  unmyelinated axons innervating the rat cranial meninges.
\newblock {\em The Journal of physiology}, 586(4):1089--1103.

\bibitem[Destexhe, 1994a]{Destexhe1994b}
Destexhe, A. (1994a).
\newblock Oscillations, complex spatiotemporal behavior, and information
  transport in networks of excitatory and inhibitory neurons.
\newblock {\em Physical Review E}, 50(2):1594.

\bibitem[Destexhe, 1994b]{Destexhe1994a}
Destexhe, A. (1994b).
\newblock Stability of periodic oscillations in a network of neurons with time
  delay.
\newblock {\em Physics Letters A}, 187(4):309--316.

\bibitem[Destexhe, 2009]{Destexhe2009}
Destexhe, A. (2009).
\newblock Self-sustained asynchronous irregular states and up--down states in
  thalamic, cortical and thalamocortical networks of nonlinear
  integrate-and-fire neurons.
\newblock {\em Journal of computational neuroscience}, 27(3):493.

\bibitem[Ernst et~al., 1995]{Ernst1995}
Ernst, U., Pawelzik, K., and Geisel, T. (1995).
\newblock Synchronization induced by temporal delays in pulse-coupled
  oscillators.
\newblock {\em Physical review letters}, 74(9):1570.

\bibitem[Gasser and Grundfest, 1939]{Gasser1939}
Gasser, H.~S. and Grundfest, H. (1939).
\newblock Axon diameters in relation to the spike dimensions and the conduction
  velocity in mammalian a fibers.
\newblock {\em American Journal of Physiology-Legacy Content}, 127(2):393--414.

\bibitem[Gautrais and Thorpe, 1998]{Gautrais1998}
Gautrais, J. and Thorpe, S. (1998).
\newblock Rate coding versus temporal order coding: a theoretical approach.
\newblock {\em Biosystems}, 48(1-3):57--65.

\bibitem[Geisler et~al., 2010]{Geisler2010}
Geisler, C., Diba, K., Pastalkova, E., Mizuseki, K., Royer, S., and
  Buzs{\'a}ki, G. (2010).
\newblock Temporal delays among place cells determine the frequency of
  population theta oscillations in the hippocampus.
\newblock {\em Proceedings of the National Academy of Sciences},
  107(17):7957--7962.

\bibitem[Gluss, 1967]{Gluss1967}
Gluss, B. (1967).
\newblock A model for neuron firing with exponential decay of potential
  resulting in diffusion equations for probability density.
\newblock {\em The Bulletin of Mathematical Biophysics}, 29(2):233--243.

\bibitem[Graupner and Reyes, 2013]{Graupner2013}
Graupner, M. and Reyes, A.~D. (2013).
\newblock Synaptic input correlations leading to membrane potential
  decorrelation of spontaneous activity in cortex.
\newblock {\em Journal of Neuroscience}, 33(38):15075--15085.

\bibitem[Harper and Lawson, 1985]{Harper1985}
Harper, A. and Lawson, S. (1985).
\newblock Conduction velocity is related to morphological cell type in rat
  dorsal root ganglion neurones.
\newblock {\em The Journal of physiology}, 359(1):31--46.

\bibitem[Hopfield, 1982]{Hopfield1982}
Hopfield, J.~J. (1982).
\newblock Neural networks and physical systems with emergent collective
  computational abilities.
\newblock {\em Proceedings of the national academy of sciences},
  79(8):2554--2558.

\bibitem[Izhikevich, 2006]{Izhikevich2006}
Izhikevich, E.~M. (2006).
\newblock Polychronization: computation with spikes.
\newblock {\em Neural Computation}, 18(2):245--282.

\bibitem[Klampfl and Maass, 2013]{Klampfl2013}
Klampfl, S. and Maass, W. (2013).
\newblock Emergence of dynamic memory traces in cortical microcircuit models
  through stdp.
\newblock {\em Journal of Neuroscience}, 33(28):11515--11529.

\bibitem[Lapicque, 1907]{Lapicque1907}
Lapicque, L. (1907).
\newblock Recherches quantitatives sur l'excitation electrique des nerfs
  traitee comme une polarization.
\newblock {\em Journal de Physiologie et de Pathologie Generalej}, 9:620--635.

\bibitem[Lee et~al., 1986]{Lee1986}
Lee, K.~H., Chung, K., Chung, J.~M., and Coggeshall, R.~E. (1986).
\newblock Correlation of cell body size, axon size, and signal conduction
  velocity for individually labelled dorsal root ganglion cells in the cat.
\newblock {\em Journal of Comparative Neurology}, 243(3):335--346.

\bibitem[Lindner, 2014]{Lindner2014}
Lindner, B. (2014).
\newblock Low-pass filtering of information in the leaky integrate-and-fire
  neuron driven by white noise.
\newblock In {\em International Conference on Theory and Application in
  Nonlinear Dynamics (ICAND 2012)}, pages 249--258. Springer.

\bibitem[Maex and De~Schutter, 2003]{Maex2003}
Maex, R. and De~Schutter, E. (2003).
\newblock Resonant synchronization in heterogeneous networks of inhibitory
  neurons.
\newblock {\em Journal of Neuroscience}, 23(33):10503--10514.

\bibitem[Mainen and Sejnowski, 1995]{Mainen1995}
Mainen, Z.~F. and Sejnowski, T.~J. (1995).
\newblock Reliability of spike timing in neocortical neurons.
\newblock {\em Science}, 268(5216):1503--1506.

\bibitem[Mehring et~al., 2003]{Mehring2003}
Mehring, C., Hehl, U., Kubo, M., Diesmann, M., and Aertsen, A. (2003).
\newblock Activity dynamics and propagation of synchronous spiking in locally
  connected random networks.
\newblock {\em Biological cybernetics}, 88(5):395--408.

\bibitem[Oppenheim et~al., 1997]{Oppenheim1997}
Oppenheim, A.~V., Willsky, A.~S., and Hamid, S. (1997).
\newblock Signals and systems.

\bibitem[Ostojic, 2014]{Ostojic2014}
Ostojic, S. (2014).
\newblock Two types of asynchronous activity in networks of excitatory and
  inhibitory spiking neurons.
\newblock {\em Nature neuroscience}, 17(4):594--600.

\bibitem[Renart et~al., 2010]{Renart2010}
Renart, A., De~La~Rocha, J., Bartho, P., Hollender, L., Parga, N., Reyes, A.,
  and Harris, K.~D. (2010).
\newblock The asynchronous state in cortical circuits.
\newblock {\em Science}, 327(5965):587--590.

\bibitem[Reyes, 2003]{Reyes2003}
Reyes, A.~D. (2003).
\newblock Synchrony-dependent propagation of firing rate in iteratively
  constructed networks in vitro.
\newblock {\em Nature neuroscience}, 6(6):593--599.

\bibitem[Smith, 2007]{Smith2007}
Smith, J.~O. (2007).
\newblock {\em Introduction to digital filters: with audio applications},
  volume~2.
\newblock Julius Smith.

\bibitem[Stein, 1965]{Stein1965}
Stein, R.~B. (1965).
\newblock A theoretical analysis of neuronal variability.
\newblock {\em Biophysical Journal}, 5(2):173--194.

\bibitem[Swadlow, 1974]{Swadlow1974}
Swadlow, H.~A. (1974).
\newblock Systematic variations in the conduction velocity of slowly conducting
  axons in the rabbit corpus callosum.
\newblock {\em Experimental neurology}, 43(2):445--451.

\bibitem[Swadlow, 1985]{Swadlow1985}
Swadlow, H.~A. (1985).
\newblock Physiological properties of individual cerebral axons studied in vivo
  for as long as one year.
\newblock {\em Journal of Neurophysiology}, 54(5):1346--1362.

\bibitem[Swadlow, 1994]{Swadlow1994}
Swadlow, H.~A. (1994).
\newblock Efferent neurons and suspected interneurons in motor cortex of the
  awake rabbit: axonal properties, sensory receptive fields, and subthreshold
  synaptic inputs.
\newblock {\em Journal of neurophysiology}, 71(2):437--453.

\bibitem[Thalhammer et~al., 1994]{Thalhammer1994}
Thalhammer, J., Raymond, S., Popitz-Bergez, F., and Strichartz, G. (1994).
\newblock Modality-dependent modulation of conduction by impulse activity in
  functionally characterized single cutaneous afferents in the rat.
\newblock {\em Somatosensory \& motor research}, 11(3):243--257.

\bibitem[Thorpe, 1990]{Thorpe1990}
Thorpe, S.~J. (1990).
\newblock Spike arrival times: A highly efficient coding scheme for neural
  networks.
\newblock {\em Parallel Processing in Neural Systems}, pages 91--94.

\bibitem[Van~Vreeswijk et~al., 1994]{vanVreeswijk1994}
Van~Vreeswijk, C., Abbott, L., and Ermentrout, G.~B. (1994).
\newblock When inhibition not excitation synchronizes neural firing.
\newblock {\em Journal of computational neuroscience}, 1(4):313--321.

\bibitem[Van~Vreeswijk and Sompolinsky, 1996]{vanVreeswijk1996}
Van~Vreeswijk, C. and Sompolinsky, H. (1996).
\newblock Chaos in neuronal networks with balanced excitatory and inhibitory
  activity.
\newblock {\em Science}, 274(5293):1724--1726.

\bibitem[Vogels and Abbott, 2009]{Vogels2009}
Vogels, T.~P. and Abbott, L. (2009).
\newblock Gating multiple signals through detailed balance of excitation and
  inhibition in spiking networks.
\newblock {\em Nature neuroscience}, 12(4):483.

\bibitem[Wang and Buzs{\'a}ki, 1996]{Wang1996}
Wang, X.-J. and Buzs{\'a}ki, G. (1996).
\newblock Gamma oscillation by synaptic inhibition in a hippocampal
  interneuronal network model.
\newblock {\em Journal of neuroscience}, 16(20):6402--6413.

\bibitem[Waxman, 1980]{Waxman1980}
Waxman, S.~G. (1980).
\newblock Determinants of conduction velocity in myelinated nerve fibers.
\newblock {\em Muscle \& Nerve: Official Journal of the American Association of
  Electrodiagnostic Medicine}, 3(2):141--150.

\bibitem[Wilson and Cowan, 1972]{Wilson1972}
Wilson, H.~R. and Cowan, J.~D. (1972).
\newblock Excitatory and inhibitory interactions in localized populations of
  model neurons.
\newblock {\em Biophysical Journal}, 12(1):1--24.

\end{thebibliography}

\section{Methods}
All simulations carried out in this paper are done with leaky intergrate-and-fire neurons \citep{Lapicque1907}, receiving $M$ correlated, weighted and time-shifted inputs $w_mI^{(m)}(t-d_m)$. The membrane potential is governed by the equation
\begin{equation}
  \tau\frac{dv}{dt} + v= G\left( I(t-d_0) + \sum_{m=1}^{M}{w_mI^{(m)}(t-d_m)}\right), \nonumber
\end{equation}
which is supplemented with a spike-and-reset mechanism: each time $v$ surpasses a threshold value $v_c$, it is said to fire a spike and is directly reset to a reset value $v_r$. Throughout a normalising gain factor of $G=(1+\sum_{m=1}^{M}|w_m|)^{-1}$ is used. In the simulations for the results and figures of this article the inputs $I^{(m)}$ to the neuron are either pure white-noise or filtered white-noise, depending on the particular simulation (see the following method sections for the specifics per simulation). In general, for simplicity, the first delay time is set to zero (i.e. $d_0 = 0$).

\subsection{Measured sub-threshold frequency response spectra}
The sub-threshold frequency response spectra are measured by driving the neuron with inputs which are time-shifted versions of a single white-noise source with mean $\mu=0$ and standard deviation $\sigma=1$. To ignore the fast membrane fluctuation caused by spiking, the spiking threshold was set to infinity ($v_c = \infty$). The simulations were carried out with $1\mathrm{e}^{4}$ timesteps per spectrum. Each spectrum is the average over the spectra of $1\mathrm{e}^{4}$ simulated neurons.

\subsection{Frequency selectivity: discrimination of inputs}
The `base' input both neurons received alternates between two different noise signals with differing spectral content, each of these signals matching the frequency selectivity of either one of the two neurons. In the first $4000$ timesteps of each realisation the administered input is an unfiltered white noise signal, in order to observe the baseline firing of each neuron. For these simulations the spike-and-reset mechanism $v>v_{th} \implies v\leftarrow v_r$ is reintroduced, so it is be possible to observe the spiking behavior of the two neurons in response to the different inputs. Simulations were carried out with $2e4$ timesteps. Each group consists of $500$ neurons, for a total of $1000$ neurons. 
The firing rates are calculated per neuron with a time-window of $200$ timesteps. The rate of each neuron is transformed into z-scores. The z-scores are averaged in each condition (white-noise, A-matched or B-matched), per neuron. Two-sample independent t-tests are carried out comparing the distribution of average z-scores of neuron A versus that of neuron B, per condition. Matched sample t-test were carried out between the firing z-scores between the random interval and the matched intervals. For plotting purposes, the firing-rate z-scores are averaged over all neurons per timestep.

\subsection{Frequency selectivity: specificity of responses}
Each neurons receives $13$ inputs with weights leading to zeros evenly spaced around the unit-circle, with the exeption of one conjugate pair. In this way the neuron is mainly responsive to a small band of frequencies around the angle of the missing pair of zeros. These neurons are subsequently exposed to bandpass filtered white noise with differing center frequencies. The spike count of each neuron in response to each bandpass filtered noise stimulus is recorded and normalised per neuron over different inputs such that the maximal count of each neuron equals one. Each pixel correspons to the average spike count of $500$ neurons over $2e3$ timesteps. Each spike rate is calculated over the whole time of each trial.

\subsection{Robustness to non perfect correlation}
Each input $I^{(m)}$ to the synapses of the neurons in these simulations is driven by an input consisting of a source input $I$, which is a specifically filtered white-noise signal matched to the prefferred spectrum of a neuron. This input is shared by all the synapses. In addition each synapse receives a unique noise $eta_m$, which is a randomly permutated version of the matched input. This input is unique to each synapse. Thus:
\begin{equation}\label{eq:rhonoise}
  I^{(m)} = \rho I + (1-\rho) \eta_m.
\end{equation}
The synapse parameters ($M$, $w_m$ and $d_m$) are contructed from $(M-1)/2$ randomly drawn zeros on the upper half of the unit circle and their conjugates, in addition to the direct input $m=0$. This resulted in $1000$ neurons with different frequency selectivity. These neurons are then driven in 6 trials with inputs with differing inter-synapse correlations from perfect correlation $\rho=1$ to completely uncorrelated $\rho=0$, as described above. Each trial consisted of 5 blocks of equal time: the first $2000$ timesteps all neurons are driven a random permutation of their matched input (shuffled along the time dimension), resulting in a white noise input. Following there are 4 blocks of $2000$ timesteps each, during the first and the third block each neuron receives their matched inputs. In the second and last block each neuron receives the matched input of another neuron, randomly drawn (thus a shuffling of the pairing between neuron and input). The firing rates are calculated over a window of $500$ timesteps, for each neuron for each correlation level, and transformed into z-scores. For each correlation level the mean firing rate z-scores of each neuron during the matched blocks is compared to its firing rate z-score during the unmachted blocks in a two-sample paired t-test. For plotting of the firing rates, the firing rate z-scores are averaged across neurons, resulting in one line per correlation level.

\end{document}